\documentclass[conference]{IEEEtran}
\usepackage{lipsum,graphicx,multicol}
\usepackage{color,soul}
\usepackage{mathtools}
\usepackage{graphicx}
\usepackage{color,soul}
\usepackage{setspace}
\usepackage{multirow}
\usepackage{tablefootnote}
\usepackage[font=small]{caption}
\usepackage{subcaption}
\usepackage{bm, amsmath, amssymb}
\usepackage{amsfonts}
\usepackage {amssymb,amsmath,xcolor}
\usepackage{epstopdf}
\usepackage{booktabs}
\usepackage{tikz}
\DeclareGraphicsExtensions{.pdf,.png,.jpg}

\ifCLASSINFOpdf
\else
\fi
\newcommand\copyrighttext{ %
	\footnotesize "This work has been submitted to IEEE PIMRC 2017 conference for possible publication. Copyright may be transferred without prior notice, after which this version may no longer be accessible"}
\newcommand\copyrightnotice{%
	\begin{tikzpicture}[remember picture,overlay]
	\node[anchor=south,yshift=760pt] at (current page.south) {{\parbox{\dimexpr\textwidth-\fboxsep-\fboxrule\relax}{\copyrighttext}}};
	\end{tikzpicture}%
}

\begin{document}
\title{Performance Optimization of Co-Existing Underlay Secondary Networks}

\author{\IEEEauthorblockN{Pratik Chakraborty}
\IEEEauthorblockA{Bharti School of Telecom. Technology and Management\\
Indian Institute of Technology Delhi\\
New Delhi-110016, India\\
Email: bsz128380@dbst.iitd.ac.in}
\and
\IEEEauthorblockN{Shankar Prakriya}
\IEEEauthorblockA{Department of Electrical Engineering and \\ Bharti School of Telecom. Technology and Management\\
Indian Institute of Technology Delhi\\
New Delhi-110016, India\\
Email: shankar@ee.iitd.ac.in}
}

\maketitle
\copyrightnotice

\begin{abstract}
	In this paper, we analyze the throughput performance of two co-existing downlink multiuser underlay secondary networks that use fixed-rate transmissions. We assume that the interference temperature limit (ITL) is apportioned to accommodate two concurrent transmissions using an interference temperature apportioning parameter so as to ensure that the overall interference to the primary receiver does not exceed the ITL. Using the derived analytical expressions for throughput, when there is only one secondary user in each network, or when the secondary networks do not employ opportunistic user selection (use round robin scheduling for example), there exists a critical fixed-rate below which sum throughput with co-existing secondary networks is higher than the throughput with a single secondary network. We derive an expression for this critical fixed-rate. Below this critical rate, we show that careful apportioning of the ITL is critical to maximizing sum throughput of the co-existing networks. We derive an expression for this apportioning parameter.  Throughput is seen to increase with increase in number of users in each of the secondary networks. Computer simulations demonstrate accuracy of the derived expressions.
\end{abstract}

\IEEEpeerreviewmaketitle

\section{Introduction}
A rapid increase in wireless devices and services  in the past decade or so has led to a demand for very high data rates over the wireless medium. With such prolific increase in data traffic, mitigating spectrum scarcity and more efficient utilization of under-utilized spectrum has drawn attention of researchers both in academia and in the industry. Cognitive radios (CR) are devices that have shown promise in alleviating these problems of spectrum scarcity and low spectrum utilization efficiencies.
\par In underlay mode of operation of cognitive radios, both secondary (unlicensed) and primary (licensed) users co-exist and transmit in parallel such that the total secondary interference caused to the primary user is below a predetermined threshold \cite{LeHossain2008} referred to as the interference temperature limit (ITL). This ensures that primary performance in terms of throughput or outage is maintained at a desired level. Most of the analysis to date in underlay CR literature is confined to one secondary node transmitting with full permissible power and catering to its own set of receivers, while maintaining service quality of the primary network. For such secondary networks, performance improvement is achieved by exploiting diversity techniques \cite{LeeWangAndrewsHong2011,YeohElkashlanKimDuongKaragiannidis2016}, resource allocation \cite{HeckeFiorentinoLotticiGiannettiVandendorpeMoeneclaey2017}, increasing the number of hops \cite{BoddapatiBhatnagarPrakriya2016}, etc. Cognitive radios have attracted research interest due to the possibility of great increase in spectrum utilization efficiency.
\par Researchers have proposed the idea  of concurrent secondary transmissions to further increase throughput (and therefore spectrum utilization efficiency), where two or more cognitive femtocells reuse the spectrum of a macrocell  either in a overlay, interweave or underlay manner \cite{ChengAoTsengChen2012}. By deploying femtocells, operators can reduce the traffic on macro base stations and also improve data quality among femtocell mobile stations due to short range communication. To  implement such an underlay scheme, the major hindrance is  mitigation of interferences among inter-femtocell users and careful handling of  interferences from femtocell transmitters to the users of the macro cell \cite{ChengLienChuChen2011}. A comprehensive survey of such heterogeneous networks, their implementation and future goals can be found in \cite{PengWangLiXiangLau2015} (and references therein).
\par  In this paper, we consider two co-existing downlink multiuser underlay networks.  We show that throughput with two co-existing secondary networks is larger than with one secondary network in some situations. Since throughput performance is ensured, this implies the possibility of increase in spectrum utilization efficiency. The main contributions of our paper are as follows:
\begin{enumerate}
	\item  Unlike other works on  co-existing secondary networks that focus on  optimization \cite{XingMathurHaleemChandramouliSubbalakshmi2007} and game theoretic approaches \cite{KangZhangMotani2012}, we present an analytical closed form sum throughput expression for two co-existing secondary multiuser downlink networks using fixed-rate transmissions by the secondary nodes.
	%	\item We address the issue of interference mitigation at the primary receiver \cite{ChengLienChuChen2011} from two secondary users by careful apportioning of the peak transmit powers and ITL such that the sum secondary interference to the primary does not exceed $I_P$.
	\item We evaluate analytically the maximum secondary fixed rate by sources that yields higher throughput with concurrent transmissions in two co-existing secondary networks. Beyond this rate,  switching to a single secondary transmission is better.
	\item We propose an optimal ITL apportioning parameter to further improve the sum throughput performance when two secondary sources transmit at the same time.
	\item We show that sum throughput improves with user selection in individual secondary networks.
\end{enumerate}
The derived expressions and insights are a useful aid to system designers. 

\section{System Model and Problem Formulation}
\par
\begin{figure}[ht]
	\begin{center}
		\includegraphics[scale = 0.5]{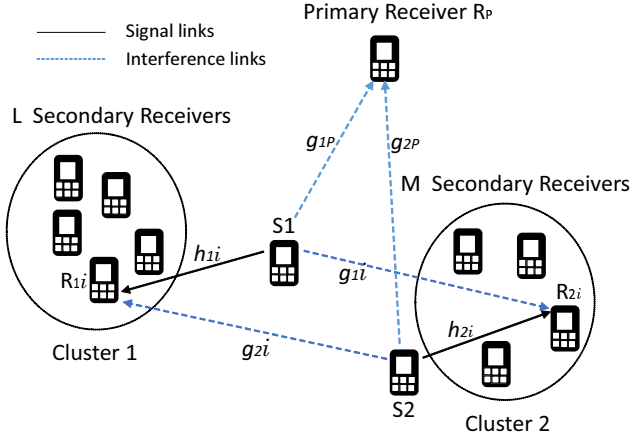}
		\caption{System model of co-existing underlay cognitive radio network}
		\label{fig:sysmod}
	\end{center}
\end{figure}

 We consider two cognitive underlay downlink networks\footnote{Although primary and secondary networks are often assumed to be licensed and unlicensed users respectively, this need not always be the case.  They can indeed be users of the same network transmitting concurrently to increase spectrum utilization efficiency. The same logic extends to two co-existing secondary networks. This eliminates most of the difficulties associated with interference channel estimation, security, etc.}, where two secondary transmitters $S_1$ and $S_2$ transmit symbols concurrently in the range of a primary network by selecting their best receivers $R_{1i^{*}}$ (among $R_{1i}$ receivers, $i \in [1,L]$) and $R_{2i^{*}}$ (among $R_{2i}$ receivers, $i \in [1,M]$) respectively, from their cluster of users (Fig. \ref{fig:sysmod}). We assume that the two secondary networks  are located relatively far apart so that the same frequency can be reused by $S_1$ and $S_2$ concurrently. We ensure that the total secondary interference caused to the primary receiver $R_{P}$ is below ITL by careful apportioning of power between $S_1$ and $S_2$.
All channels are assumed to be independent, and of  quasi-static Rayleigh fading type. The channels between $S_1$ and $R_{1i}$ are denoted by $h_{1i}\sim {\cal CN}\left(0,1/\lambda_{11}\right)$, $i \in [1,L]$. The channels between $S_2$ and $R_{2i}$ are denoted by $h_{2i}\sim {\cal CN}\left(0,1/\lambda_{22}\right)$, $i \in [1,M]$. Due to concurrent secondary transmissions, each transmitter interferes with the receivers of the other cluster. The interference channels between $S_1$ and $R_{2i}$ are denoted by $g_{1i}\sim {\cal CN}\left(0,1/\mu_{12}\right)$, $i \in [1,M]$, with $g_{1i^{*}}$ being the channel to the intended receiver $R_{2i^{*}}$. The interference channels between $S_2$ and $R_{1i}$ are denoted by $g_{2i}\sim {\cal CN}\left(0,1/\mu_{21}\right)$, $i \in [1,L]$, with $g_{2i^{*}}$ being the channel to the intended receiver $R_{1i^{*}}$. The channels to $R_{P}$ from $S_1$ and $S_2$ are denoted by $g_{1P}\sim {\cal CN}\left(0,1/\mu_{1P}\right)$ and $g_{2P}\sim {\cal CN}\left(0,1/\mu_{2P}\right)$ respectively. We neglect primary interference at the secondary nodes assuming the primary transmitter to be located far away from the secondary receivers, which is a common assumption in CR literature, and well justified on information theoretic grounds \cite{JovicicViswanath2009}, \cite{VuDevroyeTarokh2009}. Zero-mean additive white Gaussian noise of variance $\sigma_n^2$ is assumed at all terminals. As in all underlay networks, it is assumed that $S_1$ and $S_2$ can estimate $|g_{1P}|^{2}$ and $|g_{2P}|^{2}$ respectively by observing the primary reverse channel, or using pilots transmitted by $R_{P}$. 
\par In every signaling interval, $S_1$ transmits unit energy symbols $x$ with power $P_{S1} = \alpha I_P/|g_{1P}|^{2}$ and $S_2$ transmits unit energy symbols $z$ with power $P_{S2} = (1-\alpha) I_P/|g_{2P}|^{2}$, where $I_P$ denotes the ITL, and $0<\alpha<1$ denotes the power allocation parameter which apportions $I_P$ between $S_1$ and $S_2$ respectively. We use peak interference type of power control at $S_1$ and $S_2$ instead of limiting the transmit powers with a peak power due to the following reasons:\\
\begin{enumerate}
	\item It is well known that the performance of CR networks exhibits an outage floor after a certain peak power and does not improve beyond a point when transmit powers are limited by interference constraints.
	\item Since sufficient peak power is typically available, this assumption is quite reasonable. It is in this regime where cognitive radios are expected to operate. Such an assumption is also common in prior underlay CR literature \cite{DuongCostaElkashlanBao2012,TourkiQaraqeAlouini2013,ChakrabortyPrakriya2017}.
	\item It keeps the analysis tractable, leading to precise performance expressions that offer useful insights. It also allows us to derive expressions for important parameters of practical interest in the normal range of operation of secondary networks, and can yield insights of interest to system designers.
\end{enumerate}
The received signals ($y_{R_{1i}}$ and $y_{R_{2i}}$) at $R_{1i}$ and $R_{2i}$ can be written as follows:
\begin{IEEEeqnarray}{rCl}
	\nonumber y_{R_{1i}} &=& \sqrt{\frac{\alpha I_P}{|g_{1P}|^{2}}}h_{1i}x + \sqrt{\frac{(1-\alpha) I_P}{|g_{2P}|^{2}}}g_{2i}z + n_{R_{1i}}, i \in [1,L]\\
	\nonumber y_{R_{2i}} &=& \sqrt{\frac{(1-\alpha) I_P}{|g_{2P}|^{2}}}h_{2i}z + \sqrt{\frac{\alpha I_P}{|g_{1P}|^{2}}}g_{1i}x + n_{R_{2i}}, i \in [1,M],\\
	\label{eq: signals}
\end{IEEEeqnarray}
where $n_{R_{1i}},n_{R_{2i}}\sim {\cal CN}(0,\sigma^{2}_{n})$ are additive white Gaussian noise samples at $R_{1i}$ and $R_{2i}$ respectively. When transmitters $S_1$ and $S_2$ select the receivers $R_{1i^{*}}$ and $R_{2i^{*}}$ with strongest link to them in their individual cluster, the instantaneous signal-to-interference-plus-noise ratios (SINRs) $\Gamma_{1}$ and $\Gamma_{2}$ at $R_{1i^{*}}$ and $R_{2i^{*}}$ can be written as  follows:\\
\begin{IEEEeqnarray}{rCl}
	\nonumber \Gamma_{1} &=& \frac{\alpha I_P \frac{\displaystyle \max_{i \in [1,L]}{[|h_{1i}|^{2}]}}{|g_{1P}|^{2}} }{(1-\alpha) I_P \frac{|g_{2i^{*}}|^{2}}{|g_{2P}|^{2}} + \sigma_n^2}\\
	\Gamma_{2} &=& \frac{(1-\alpha) I_P \frac{\displaystyle \max_{i \in [1,M]}{[|h_{2i}|^{2}]}}{|g_{2P}|^{2}} }{\alpha I_P \frac{|g_{1i^{*}}|^{2}}{|g_{1P}|^{2}} + \sigma_n^2}.
	\label{eq: SINRS_main}
\end{IEEEeqnarray}
We note that the random variables $|h_{ij}|^2$ and $|g_{ij}|^2$ in \eqref{eq: SINRS_main} follow the exponential distribution with mean values $1/\lambda_{ii}$ and $1/\mu_{ij}$ respectively.
\par In the following section, we derive sum throughput expression for this co-existing secondary network. It gives a measure of spectrum utilization with or without concurrent transmissions by sources in co-existing secondary networks.
\section{Sum Throughput of the Secondary Network for Fixed Rate Transmission Scheme}
\par When secondary nodes transmit with a fixed rate $R$,  the sum throughput $\tau_{sum}$ is given by:
\begin{IEEEeqnarray}{rCl}
	\tau_{sum} &=& (1-p_{out_1})R + (1-p_{out_2})R,
	\label{eq: sum_th_fixed_rate_def}
\end{IEEEeqnarray}
where $p_{out_1}$ and $p_{out_2}$ are outage probabilities of the two secondary user pairs $S_1$-$R_{1i^{*}}$ and $S_2$-$R_{2i^{*}}$ respectively.
\subsection{Derivation of $p_{out_1}$:}
The outage probability $p_{out_1}$ is defined as follows:
\begin{IEEEeqnarray}{rCl}
	\nonumber p_{out_1} &=& \Pr\{\Gamma_1 < \gamma_{th}\},
	\label{eq: p_OS1_def}
\end{IEEEeqnarray}
where $\gamma_{th} = 2^{R}-1$. For notational convenience, we define random variable $X = \displaystyle \max_{i \in [1,L]}[|h_{1i}|^2]$. Clearly, it has cumulative distribution function (CDF) $F_X(x) = (1 - e^{\lambda_{11}x})^L$. Thus, $p_{out_1}$ can be rewritten and evaluated as under:
\begin{IEEEeqnarray}{rCl}
	\nonumber p_{out_1} &=& \Pr \bigg\{X \hspace{-0.1 cm} < \hspace{-0.1 cm} \bigg(\frac{1-\alpha}{\alpha}\bigg)\gamma_{th}\frac{|g_{1P}|^2}{|g_{2P}|^2}|g_{2i^{*}}|^2 + \frac{\gamma_{th}\sigma_n^2}{\alpha I_P}|g_{1P}|^2 \bigg\}
\end{IEEEeqnarray}
\begin{IEEEeqnarray}{rCl}
	\nonumber &=& \mathbb{E} \bigg[ \bigg( 1 - e^{-\lambda_{11}\{(\frac{1-\alpha}{\alpha})\gamma_{th}\frac{|g_{1P}|^2}{|g_{2P}|^2}|g_{2i^{*}}|^2 + \frac{\gamma_{th}\sigma_n^2}{\alpha I_P}|g_{1P}|^2\}}  \bigg)^L ~ \bigg]\\
	\nonumber &=& \mathbb{E}\bigg[1 - \sum\limits_{j=1}^{L}\dbinom{L}{j}(-1)^{j+1}\\
	&& \hspace{1.5 cm} e^{-\lambda_{11}j\{(\frac{1-\alpha}{\alpha})\gamma_{th}\frac{|g_{1P}|^2}{|g_{2P}|^2}|g_{2i^{*}}|^2 + \frac{\gamma_{th}\sigma_n^2}{\alpha I_P}|g_{1P}|^2\}} \bigg],
	\label{eq: p_OS1_1}
\end{IEEEeqnarray}
where $\mathbb{E}[.]$ denotes the expectation over random variables $|g_{1P}|^2$, $|g_{2P}|^2$ and $|g_{2i^{*}}|^2$. We evaluate $p_{out_1}$ by successive averaging over random variables $|g_{2i^{*}}|^2$, $|g_{2P}|^2$ and $|g_{1P}|^2$ using standard integrals \cite[eq.(3.353.5)]{gradshteyn2007} and \cite[eq.(4.2.17)]{geller1969table}. A final closed form expression for $p_{out_1}$ can be derived as follows (details omitted due to space limitations): 
\begin{IEEEeqnarray}{rCl}
	\nonumber p_{out_1} &=& 1 - \sum\limits_{j=1}^{L}\dbinom{L}{j}(-1)^{j+1} \bigg[ \frac{1}{1+\frac{\lambda_{11}j\gamma_{th}\sigma_n^2}{\mu_{1P}\alpha I_P}}\\
	\nonumber &&  -\frac{\frac{\mu_{2P}\lambda_{11}}{\mu_{1P}\mu_{21}}j(\frac{1-\alpha}{\alpha})\gamma_{th} \bigg\{ \ln\bigg(\frac{1 + \frac{\lambda_{11}j\gamma_{th}\sigma_n^2}{\mu_{1P}\alpha I_P}}{\frac{\mu_{2P}\lambda_{11}}{\mu_{1P} \mu_{21}}j(\frac{1-\alpha}{\alpha})\gamma_{th}}\bigg)}{[1 + \frac{\lambda_{11}j\gamma_{th}\sigma_n^2}{\mu_{1P}\alpha I_P} - \frac{\mu_{2P}\lambda_{11}}{\mu_{1P} \mu_{21}}j(\frac{1-\alpha}{\alpha})\gamma_{th}]^2}\\
	&& \hspace{0.6in}+  \bigg(\frac{\frac{\mu_{2P}\lambda_{11}}{\mu_{1P} \mu_{21}}j(\frac{1-\alpha}{\alpha})\gamma_{th}}{1 + \frac{\lambda_{11}j\gamma_{th}\sigma_n^2}{\mu_{1P}\alpha I_P}}\bigg) - 1 \bigg\}  \bigg].
	\label{eq: p_OS1_final}
\end{IEEEeqnarray}
\subsection{Derivation of $p_{out_2}$:}
The outage probability $p_{out_2}$ is defined as follows:
\begin{IEEEeqnarray}{rCl}
	p_{out_2} &=& \Pr\{\Gamma_2 < \gamma_{th}\}.
	\label{eq: p_OS2_def}
\end{IEEEeqnarray}
Due to the identical nature of SINR-s of $\Gamma_1$ and $\Gamma_2$, $p_{out_2}$ in \eqref{eq: p_OS2_def} can be derived in the same manner as $p_{out_1}$, whose final closed form expression is shown as follows:
\begin{IEEEeqnarray}{rCl}
	\nonumber p_{out_2} &=& 1 - \sum\limits_{k=1}^{M}\dbinom{M}{k}(-1)^{k+1} \bigg[ \frac{1}{1+\frac{\lambda_{22}k\gamma_{th}\sigma_n^2}{\mu_{2P}(1-\alpha) I_P}}\\
	\nonumber &&  -\frac{\frac{\mu_{1P}\lambda_{22}}{\mu_{2P}\mu_{12}}k(\frac{\alpha}{1-\alpha})\gamma_{th} \bigg\{ \ln\bigg(\frac{1 + \frac{\lambda_{22}k\gamma_{th}\sigma_n^2}{\mu_{2P}(1-\alpha) I_P}}{\frac{\mu_{1P}\lambda_{22}}{\mu_{2P} \mu_{12}}k(\frac{\alpha}{1-\alpha})\gamma_{th}}\bigg)}{[1 + \frac{\lambda_{22}k\gamma_{th}\sigma_n^2}{\mu_{2P}(1-\alpha) I_P} - \frac{\mu_{1P}\lambda_{22}}{\mu_{2P} \mu_{12}}k(\frac{\alpha}{1-\alpha})\gamma_{th}]^2}\\
	&& \hspace{0.6in} +  \bigg(\frac{\frac{\mu_{1P}\lambda_{22}}{\mu_{2P} \mu_{12}}k(\frac{\alpha}{1-\alpha})\gamma_{th}}{1 + \frac{\lambda_{22}k\gamma_{th}\sigma_n^2}{\mu_{2P}(1-\alpha) I_P}}\bigg)- 1 \bigg\}  \bigg].
	\label{eq: p_OS2_final}
\end{IEEEeqnarray}
\section{Optimal Power Allocation and Critical Target Rate}
Our objective is to find the optimum $\alpha$ (denoted by $\alpha^{*}$) that maximizes $\tau_{sum}$. From \eqref{eq: sum_th_fixed_rate_def}, it is clear that $\alpha^{*} = \arg \displaystyle \max_{\alpha}(\tau_{sum})$. In normal mode of operation, the interference channel variances are small ($\mu_{1P}$ and $\mu_{2P}$ are large) so that  $\lambda_{11}<<\mu_{1P}I_P$ and $\lambda_{22}<<\mu_{2P}I_P$. Hence, the terms $\frac{\lambda_{11}j\gamma_{th}\sigma_n^2}{\mu_{1P}\alpha I_P}$ and $\frac{\lambda_{22}k\gamma_{th}\sigma_n^2}{\mu_{2P}(1-\alpha) I_P}$ in \eqref{eq: p_OS1_final} and \eqref{eq: p_OS2_final} respectively are small quantities for practical values of target rates and can be ignored. (Computing $\alpha^{*}$ for high target rates is not required, as would become apparent in subsequent discussions.)
%Since $\tau_{sum}^{P\rightarrow \infty}$ is insensitive to variation in $I_P/\sigma_n^2$ and attains saturation for practical $I_P/\sigma_n^2$ values (Fig. 3),
Thus $p_{out_1}$ and $p_{out_2}$ reduce to the following form with $x = \frac{\mu_{2P}\lambda_{11}}{\mu_{1P}\mu_{21}}(\frac{1-\alpha}{\alpha})$ and $y = \frac{\mu_{1P}\lambda_{22}}{\mu_{2P}\mu_{12}}(\frac{\alpha}{1-\alpha})$:
\begin{IEEEeqnarray}{rCl}
	\nonumber p_{out_1} &\approx& \sum\limits_{j=1}^{L}\dbinom{L}{j}(-1)^{j+1}\gamma_{th} xj\frac{(\gamma_{th}xj - \ln(\gamma_{th}xj) - 1)}{(1-\gamma_{th}xj)^2},\\
	\nonumber && \hspace{-1.25 cm} p_{out_2} \approx \sum\limits_{k=1}^{M}\dbinom{M}{k}(-1)^{k+1}\gamma_{th} yk\frac{(\gamma_{th}yk - \ln(\gamma_{th}yk) - 1)}{(1-\gamma_{th}yk)^2}.\\
	\label{eq: IP_approx_asymptote}
\end{IEEEeqnarray}
Using the first order rational approximation for logarithm \cite{Topsoe} $\ln(z)\approx \frac{2(z-1)}{(z+1)}$ in \eqref{eq: IP_approx_asymptote}, which is close to (or follows) the logarithm function for a large range of $z$ (and also used in underlay literature \cite{ChakrabortyPrakriya2017}), $z\frac{(z-\ln(z)-1)}{(1-z)^2}\approx\frac{z}{z+1}$. Hence, $p_{out_i}, i \in \{1,2\}$ in \eqref{eq: IP_approx_asymptote} can further be approximated as:
\begin{IEEEeqnarray}{rCl}
	\nonumber p_{out_1}  &\approx& 1 - \sum\limits_{j=1}^{L}\dbinom{L}{j}(-1)^{j+1}\frac{1}{\gamma_{th}xj+1},\\
	&& \hspace{-1.25 cm} p_{out_2} \approx 1- \sum\limits_{k=1}^{M}\dbinom{M}{k}(-1)^{k+1}\frac{1}{\gamma_{th}yk+1}.
	\label{eq: IP_and_log_approx_asymptote}
\end{IEEEeqnarray}
Obtaining $\alpha^{*}$ for general $L$ and $M$ is mathematically tedious, and can be evaluated offline by numerical search\footnote{We note that there is no dependence on instantaneous channel estimates.}. However, we present a closed form $\alpha^{*}$ for the special case  when $L=M=1$.  By taking the first derivative of $\tau_{sum}$ with respect to $\alpha$ using $p_{out_1}$ and $p_{out_2}$ in \eqref{eq: IP_and_log_approx_asymptote}, and equating it to zero, a closed form $\alpha^{*}$ can be obtained\footnote{We will present a detailed proof in the extended version of this paper.}  with the root in [0,1] being:
\begin{IEEEeqnarray}{rCl}
	\alpha^{*} &\approx& \frac{1}{1 + \frac{\mu_{1P}}{\mu_{2P}} \sqrt{\frac{\lambda_{22}}{\lambda_{11}}\frac{\mu_{21}}{\mu_{12}}}}.
	\label{eq: alpha_opt}
\end{IEEEeqnarray}
\par By taking the second derivative of $\tau_{sum}$ with respect to $\alpha$, and upon substitution of $\alpha^{*}$ from \eqref{eq: alpha_opt}, an expression is obtained, which can either be positive or negative depending on the value of $\gamma_{th}$ (details are omitted due to space constraints). By equating the expression to zero and solving for $\gamma_{th}$ (or equivalently for $R$), a closed form expression of critical target rate $R=R_{c}$ (for $L=M=1$) can be obtained$^3$ as:
\begin{IEEEeqnarray}{rCl}
	R_{c} &\approx& \log_{2}\bigg(1 + \sqrt{\frac{\mu_{12}\mu_{21}}{\lambda_{11}\lambda_{22}}}\bigg).
	\label{eq: critical_rate}
\end{IEEEeqnarray}
When $R<R_c$, $\tau_{sum}$ is concave with respect to $\alpha$ and concurrent transmission offers higher throughput. When $R>R_c$, switching to single secondary transmission is optimal, as $\tau_{sum}$ is convex with respect to $\alpha$. For a generalized $L$ and $M$ users, $R_c$ and $\alpha^{*}$ can be evaluated by an offline numerical search$^2$.
%We note, as the network has a symmetrical structure, $\alpha^{*}$ is always the extrema and remains constant for a given system parameters irrespective of $\tau_{sum}^{P \rightarrow \infty}$ being concave or convex.

\par For larger $L$ and $M$ (multiple secondary users in each network), when a round robin scheduling scheme is used, the channel characteristics are exponential (same as when $L=M=1$), and (\ref{eq: alpha_opt}) and (\ref{eq: critical_rate}) are valid for $\alpha^{*}$ and $R_{c}$. We emphasize that $R_c$ and $\alpha^{*}$ both depend only on statistical channel parameters and do not require real-time computation.
\section{Simulation Results}
In this section, we present simulation results to validate the derived expressions and bring out useful insights. We assume $\mathbb{E}[|h_{ij}|^2] \propto d_{ii}^{-\phi}$, $d_{ii}$ being the normalized distance between the transmitter and intended receiver in cluster $i$, where $i \in \{1,2\}$ and $j \in \{L,M\}$. Again, $\mathbb{E}[|g_{ij}|^2] \propto r_{ij}^{-\phi}$ is assumed, where $r_{ij}$ is the normalized distance between the transmitter of cluster $i$ to the receiver of cluster $j$, where, $i \in \{1,2\}$ and $j \in \{L,M,P\}$. The pass-loss exponent is denoted by $\phi$ (assumed to be $3$ in this paper).
%\begin{figure*}[t]
%	\begin{multicols}{3}
%		\includegraphics[height=5cm, width=0.34\textwidth]{sum_th_vs_alpha_diff_rates.eps}\par\caption{$\tau_{sum}$ vs $\alpha$ for $R = 1, 2, R_c, 5$.} 
%		\includegraphics[height=5cm, width=0.35\textwidth]{sum_th_vs_alpha_diff_channels.eps}\par\caption{$\tau_{sum}$ vs $\alpha$ with optimum power allocation parameters.}
%		\includegraphics[height=5cm,width=0.34\textwidth]{sum_th_vs_rate_for_increasing_L_M.eps}\par\caption{$\tau_{sum}$ vs $R$ with $L = M = 1, 3, 5, 7, 10$.}
%	\end{multicols}
%\end{figure*}
\begin{figure}[ht]
	\begin{center}
		\includegraphics[scale = 0.3]{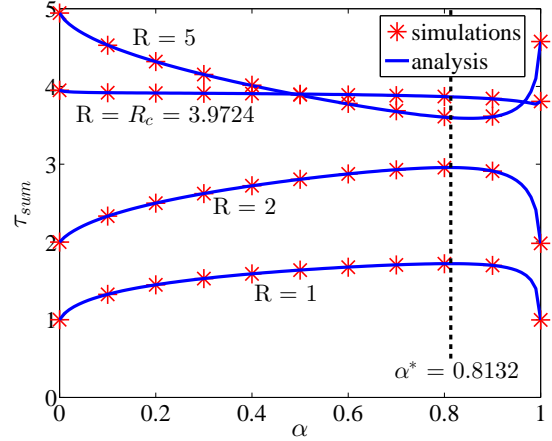}
		\caption{$\tau_{sum}$ vs $\alpha$ for $R = 1, 2, R_c, 5$.}
		\label{fig:fig1}
	\end{center}
\end{figure}
\begin{figure}[ht]
	\begin{center}
		\includegraphics[scale = 0.3]{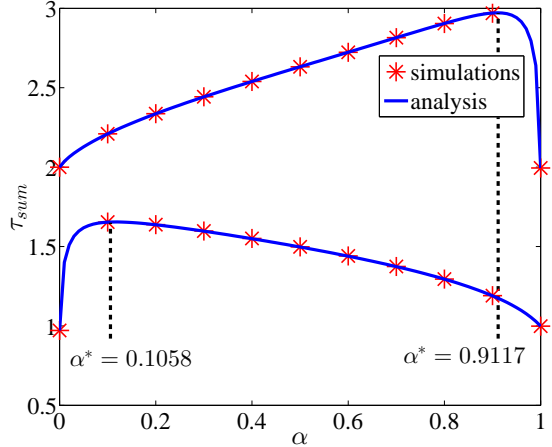}
		\caption{$\tau_{sum}$ vs $\alpha$ with optimum power allocation for different channel parameters, target rates and ITL}
		\label{fig:fig2}
	\end{center}
\end{figure}
\begin{figure}[ht]
	\begin{center}
		\includegraphics[scale = 0.3]{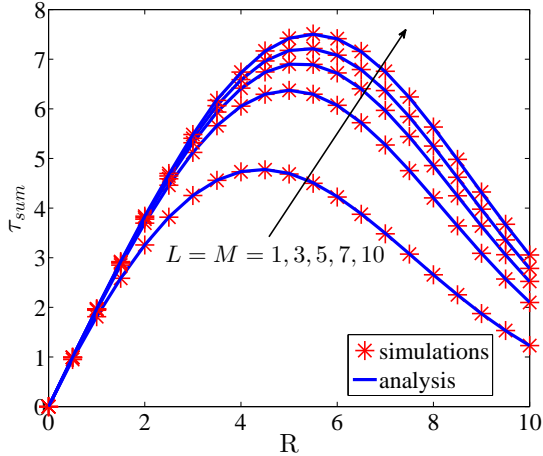}
		\caption{$\tau_{sum}$ vs $R$ with $L = M = 1, 3, 5, 7, 10$.}
		\label{fig:fig3}
	\end{center}
\end{figure}
\par In Fig. 2 we plot $\tau_{sum}$ vs $\alpha$ for different target rates. The system parameters chosen are as follows: $d_{11} = 2$ units, $d_{22} = 1$ unit, $r_{1P} = r_{2P} = 3$ units, $r_{12} = 4$ units, $r_{21} = 3$ units. $L=M=1$ and $I_P = 20 dB$ is assumed. When target rates are below $R_c=3.9724$ (as calculated from \eqref{eq: critical_rate}), there is an improvement in sum throughput of the order of $1$ bpcu when optimum $\alpha$ is chosen using concurrent transmission. If $R$ exceeds $R_c$, switching to single secondary network is best. This happens because with high target rates, both user pairs suffer link outages, and mutual interferences further degrades performance. Switching to a single network not only improves transmit power, but also nullifies the interference from the other network, which cumulatively improve outage and throughput performance.
%However, it is worth mentioning that $R_c$ as proposed in \eqref{eq: critical_rate} is valid only when the two secondary networks operate in presence of mutual interferences. When they are sufficiently far apart such that mutual interferences are almost negligible (possible when $\mu_{12}, \mu_{21} \rightarrow \infty$), there will be an influence of primary interference constraints on $R_c$.
\par In Fig. 3 we plot $\tau_{sum}$ vs $\alpha$ assuming $L=M=1$ for varying channel parameters, target rates and ITL to show that $\alpha^{*}$ as evaluated in \eqref{eq: alpha_opt} gives a fairly accurate and robust measure of optimal ITL apportioning between $S_1$ and $S_2$, and  improves sum throughput performance. The system parameters chosen for the first plot are as follows: $d_{11} = 1$ unit, $d_{22} = 2$ units, $r_{1P} = 4$ units, $r_{2P} = 3$ units, $r_{12} = 3$ units, $r_{21} = 3.5$ units and $I_P$ is chosen as $10 dB$. $R=1$ is assumed to ensure that $R<R_c=3.7037$ (so that concurrent transmission is advantageous). $\alpha^{*}=0.1058$ is obtained from \eqref{eq: alpha_opt}. In the second plot, we assume the following parameters: $d_{11} = 2$ unit, $d_{22} = 1$ units, $r_{1P} = 3$ units, $r_{2P} = 4$ units, $r_{12} = 4$ units, $r_{21} = 3$ units and $I_P$ is chosen as $25 dB$. $R=2$ is assumed to ensure that $R<R_c=3.9724$ (so that concurrent transmission is advantageous). $\alpha^{*}=0.9117$ is obtained from \eqref{eq: alpha_opt}. We note, for symmetric channel conditions, $ie$ $\lambda_{11}=\lambda_{22}$, $\mu_{12}=\mu_{21}$ and $\mu_{1P}=\mu_{2P}$, $\alpha^{*}=0.5$, implying equal resource allocation between $S_1$ and $S_2$. In addition we have the following observations: 1) $\alpha$ decreases when the ratio $\frac{\mu_{1P}}{\mu_{2P}}$ increases, or when $S_2$ is closer to the primary than $S_1$. This implies throughput can be maximized if more power is allocated to $S_2$ (thereby improving its outage), as $S_1$ has a weaker channel to primary (has more available power) and can meet its outage requirement with less transmit power. 2) $\alpha$ decreases with increase in $\frac{\lambda_{22}}{\lambda_{11}}$. In other words, when $S_1$-$R_{1i^{*}}$ channel is better than $S_2$-$R_{2i^{*}}$, $S_1$ is able to meet its outage requirement with less power, and more power needs to be allocated to $S_2$ to improve performance. 3) $\alpha$ decreases with the ratio $\frac{\mu_{21}}{\mu_{12}}$, or when the channel between $S_1$ to $R_{2i^{*}}$ is better than the channel between $S_2$ to $R_{1i^{*}}$. Thus, allocating more power to $S_2$ causes less interference to users of $S_1$, which improves the overall throughput.
\par In Fig. 4, we plot $\tau_{sum}$ in \eqref{eq: sum_th_fixed_rate_def} vs $R$ and show the effect of number of users in the two networks  on sum throughput performance with concurrent transmissions. We choose parameters as follows: $d_{11} = d_{22} = 1$ unit, $r_{1P} = r_{2P} = 3$ units and $r_{12} = r_{21} = 3$ units. $\alpha=0.5$ and $I_P = 20 dB$ is chosen. Clearly,  $\tau_{sum}$ increases with $L$ and $M$. From \eqref{eq: critical_rate}, it is also clear that $R_c$ increases with user selection (this $R_c$ refers to a network having generalized $L$ and $M$ users, which is not derived in this paper. However, intuitively it is clear that user selection statistically improves the main channels, thereby increasing $R_c$ as in \eqref{eq: critical_rate}), which causes a rightward shift of the peaks of $\tau_{sum}$. As also evident from earlier discussions, $\tau_{sum}$ first increases and then decreases after a certain critical rate as both $S_1$-$R_{1i^{*}}$ and $S_2$-$R_{2i^{*}}$ links start to suffer from outages, thereby decreasing the overall throughput performance with concurrent transmissions.
%\par \textcolor{blue}{However, computing the absolute maxima of $\tau_{sum}$ is a joint optimization problem with respect to $\alpha$ and $R$. As both are statistical quantities, such an optimization can be done offline. Finding an optimum closed form for both parameters can be a subject of future research.}
\section{Conclusion}
In this paper we analyze the sum throughput performance of two co-existing underlay multiuser secondary downlink networks utilizing fixed-rate transmissions. In the single user scenario, or in a multiuser scenario without opportunistic user selection, we establish that there exists a fixed critical rate beyond which co-existing secondary networks results in lower throughput. During concurrent secondary transmissions, we establish that user selection as well as judicious interference temperature apportioning, can increase throughput performance.
\section{Acknowledgment}
This  work  was  supported  by  Information  Technology  Research  Academy
through sponsored project ITRA/15(63)/Mobile/MBSSCRN/01. The authors thank Dr. Chinmoy Kundu for his inputs on this work.
\bibliographystyle{IEEEtran}
\bibliography{IEEEabrv,references_lit}
\end{document}